# Transdisciplinary research: How much is academia heeding the call to work more closely with societal stakeholders such as industry, government, and nonprofits?

Philip J. Purnell

Centre for Science and Technology Studies,

Leiden University,

P.O. Box 905, 2300 AX Leiden, The Netherlands

p.j.purnell@cwts.leidenuniv.nl

ORCID: 0000-0003-3146-2737

## Abstract

Transdisciplinary research, the co-creation of scientific knowledge by multiple stakeholders, is considered essential for addressing major societal problems. Research policy makers and academic leaders frequently call for closer collaboration between academia and societal stakeholders to address the grand challenges of our time. This bibliometric study evaluates progress in collaboration between academia and three societal stakeholders: industry, government, and nonprofit organisations. It analyses the level of co-publishing between academia and these societal stakeholders over the period 2013-2022. We found that research collaboration between academia and all stakeholder types studied grew in absolute terms. However, academia – industry collaboration declined 16% relative to overall academic output while academia – government and academia – nonprofit collaboration grew at roughly the same pace as academic output. Country and field of research breakdowns revealed wide variance. In light of previous work, we consider potential explanations for the gap between policymakers' aspirations and the real global trends. This study is a useful demonstration of large-scale, quantitative bibliometric techniques for research policymakers to track the impact of decisions related to funding, intellectual property law, and nonprofit support.

## Keywords

Transdisciplinary research — research stakeholders – industry – government – nonprofits

## 1. Introduction

The co-creation of knowledge focusing on complex social problems by multiple stakeholders has become known as transdisciplinary research (Hernandez-Aguilar et al., 2020) and is ideally suited to address issues that transcend scientific disciplines (Hirsch Hadorn et al., 2006). Research policymakers have recognised the potential of transdisciplinary research in tackling complicated societal challenges and frequently advocate greater research cooperation between academia and industry (Banal-Estañol et al., 2015; National Institutes of Health, 2012, 2013) citing important contributions to the economy and demonstrable benefits to all stakeholders (Lee, 2000).

Academic leaders recognise this potential and frequently call for closer collaboration with industry to bring together the 'cleverest scientists and the smartest entrepreneurs' (e.g., The Russell Group of Universities, 2023) and motivate their own academics to pursue partnerships (e.g., ETH Zürich, 2024). Large, prestigious academic institutes see such collaborations as markers of success (e.g., Tsinghua University, 2024), and appeal directly to industry stakeholders to initiate partnership proposals (e.g., Stanford University, 2024).

In parallel, collaboration between academia and government is seen as an effective method of transferring subject matter expertise to policymakers through appointing academics onto advisory boards or hiring them into dual employment positions (Glied et al., 2018; Haddon & Sasse, 2019). Meanwhile nonprofit organisations are increasingly working in partnership in sometimes chaotic, but nevertheless crucial partnerships with academia and other public bodies to address complex social, environmental, and economic problems (Cornforth et al., 2014). In each of these tandem arrangements, the potential of transdisciplinary research to address complex challenges facing communities is the driver behind the collaboration.

When the three stakeholders: academia, industry, and government work together, the collaboration is known as the 'triple helix' (Etzkowitz & Leydesdorff, 1995; Leydesdorff, 2003). The resulting inter-stakeholder co-creation of knowledge is said to offer benefits over the same sectors working in isolation (Etzkowitz & Leydesdorff, 2000) and has been proposed as an effective method of addressing the grand challenges of our time. The model has been extended to include representatives of civic society such as nonprofit organisations and the configuration of all four stakeholders working together is known as the quadruple helix (Arnkil et al., 2010).

Despite all the calls for a shift towards transdisciplinary research and inter-stakeholder collaboration, there is little evidence of progress on a global scale. Much of the literature on transdisciplinary research has focussed on the concept, definition, and the precise relationship between the stakeholders using case studies rather than the resulting research output (Carayannis et al., 2014; Marijan & Sen, 2022). Some studies have quantified papers that include the term 'transdisciplinary research' (Hernandez-Aguilar et al., 2020), however, the mere presence of the term does not mean the paper represents transdisciplinary research as defined as including contributors from multiple research stakeholders. Papers that mention transdisciplinary research are therefore not necessarily real examples of transdisciplinary research.

This study aims to use determine the extent of progress in academic collaboration with three societal stakeholders: industry, nonprofit organisations, and government over a 10-year period. Specifically, we will quantify the increase in collaborative publications in both absolute terms and as a share of the overall academic output. We expect the findings will serve as useful feedback that will help policymakers determine the success of their calls for greater inter-stakeholder collaboration. We will also conduct a breakdown of results for each of the collaboration partners by country and by field to see whether any region-specific or field-specific lessons can be learned.

## 2. Literature review

### 2.1. Modes of research and transdisciplinarity

The application of transdisciplinarity to scientific research accelerated in the 1990s in response to a perceived disconnect between the scientific community and real-world problems (Hernandez-Aguilar et al., 2020), "The world has problems, but universities have departments" (Brewer, 1999). Publication of *The new production of knowledge* by (Gibbons et al., 1994) was a key moment because the authors described the distinction between two types of research; traditional 'Mode 1' research in which universities behave autonomously conducting research in strict disciplines and isolation from society, and 'Mode 2' knowledge production through transdisciplinary collaboration by temporary research teams comprising experts with diverse backgrounds and knowledge.

Transdisciplinarity can be therefore considered a characteristic of mode 2 research (Gibbons et al., 1994) and its definition is the subject of much debate. Indeed, the lack of a universally accepted definition may offer an opportunity for evolution of the term and practice with the participation of stakeholders themselves. A longitudinal study is currently under way to assess the perspectives of diverse actors in a transdisciplinary project (Thompson et al., 2017) which may contribute to the debate. General consensus of the transdisciplinary concept describes scientists working closely with other society stakeholders to conceive, conduct, and publish research (Mauser et al., 2013), thereby co-creating knowledge with society, rather than for society (Greenhalgh et al., 2016; Seidl et al., 2013). Solutions to the grand challenges of our time such as those described by the UN sustainable development goals (SDGs) (United Nations General Assembly, 2015) will require collaborative approaches involving diverse stakeholders and decision makers (Hernandez-Aguilar et al., 2020; Thompson et al., 2017).

## 2.2. Inter-stakeholder collaboration

Many have suggested that university – industry collaboration is good for innovation and the economy (Savage, 2017) although motivation for industry scientists may be different from that of academics. Some argue that academics who collaborate with industry researchers may experience a decline in publication rates because of the company's interest in protecting research results through patents rather than publishing it. Czarnitzki et al. (2015) showed that scientists in Germany experience more than twice as much delay and secrecy, jeopardising publication disclosure of academic research when projects use industry sponsorship. Perkmann & Walsh (2009) argued however, the type of collaboration makes a difference and that more basic collaborative projects between university and company scientists results in increased publication.

The difference between academic and industry scientists may result from early career choices defined by underlying preferences and professional goals. It has been suggested that scientists driven by a 'taste for science' will more likely opt for a career in academia (Agarwal & Ohyama, 2012) while those who choose industry are more motivated by financial and other resource incentives along with greater interest in downstream research (Roach & Sauermann, 2010).

Bikard et al. (2018) observed simultaneous discoveries and found that those resulting from academia – industry partnerships were followed by effective division of follow-on work with academics dedicated to publications while company scientists concentrated on protecting intellectual property. University – industry research collaboration has been found helpful for university lecturers provided it doesn't distract them from their main academic research (Manjarrés-Henríquez et al., 2009).

Various studies (e.g., Tijssen et al., 2016) have shown geographical proximity to correlate with university-industry-government collaboration (Ponds et al., 2007). The closer the company is to the university, the more likely collaboration occurs, and research is published (Bjerregaard, 2010). In 2009, (Tijssen et al., 2009) estimated that 7% of scholarly publications featured both university and industry affiliated authors.

Since the 1990s, the Chinese government has invested consistently in research-intensive universities (Zhang et al., 2013) and at the same time, private Chinese firms have boosted their investment in R&D more than state-owned enterprises (Fang et al., 2017). In 2015, the Chinese government implemented a series of laws and policies designed to enable universities to decide their own technology transfer strategy, which may have stimulated academia – industry partnerships. Academic institutions' intellectual property is considered state-owned property, and its potential improper handling was previously seen as a deterrent to industry collaboration (Zhang & Zou, 2022).

In a US-China comparison, Zhou et al. (2016) found that Chinese universities are more likely than US universities to co-author papers with foreign industry partners. Regardless of the motivation and structure of the partnership, there seems no doubt that academia – industry collaboration is being encouraged and incentivised around the world.

Nonprofit organisations are increasing held accountable through performance assessment (Lee & Nowell, 2014), which usually encompasses financial indicators (Kim, 2016), but not output metrics, which are considered too simplistic (Rey-Garcia et al., 2017). In the book *net positive* (Polman & Winston, 2021), former Unilever CEO Paul Polman provides an entertaining personal account of steering a multinational company through turbulent times. In his view, it is vital for successful businesses to work hand in hand with government, and civic society to ensure mutual benefit, thereby prioritizing stakeholder value over shareholder value. Polman sees potential for businesses to address social challenges but emphasizes the need to do so in cooperation with other stakeholders including governments, nonprofits, and academia. As our challenges become more urgent and interdependent, we need to think of new ways of addressing them. Collaboration with nonprofit organisations will be key to finding solutions and we cannot expect governments, business, or academia to solve these challenges alone.

Government is the predominant source of research funding for the academic sector in most countries, and major changes to research funding policy influence academic publication rates, intellectual property protection, and other innovation indicators (Adams et al., 2005; Beaudry & Allaoui, 2012). The academic – government research relationship is therefore at least partially based on academics' use of government funding grants to conduct research related to its area of interest. Government research grants are sometimes the result of progammatic, mission-oriented agencies that have a need to demonstrate immediately useful research findings to achieve specific goals that do not necessarily fit traditional academic research objectives (Goldfarb, 2008). Data collected by China's Ministry of Science and Technology in 2020 showed that government – academia partnerships have been more successful than other collaboration configurations at business incubation of new R&D institutions (Zhou & Wang, 2023).

In many industrialised countries, the share of university research funding directly supported through government grants is falling while industry support is increasing (Auranen & Nieminen, 2010). In the UK, the government created UK Research and Innovation (UKRI) with the goal of unifying the voice of academia and improving cooperation with government policymakers (Haddon & Sasse, 2019). Germany spent 120 billion Euros on R&D in 2022, twice as much as France and amounting to more than one-third of the entire EU investment, while Switzerland invests 50% more in R&D than the EU average as a share of its GDP (Eurostat, 2024). Government subsidisation of R&D cooperation has stimulated innovation efficiency in Germany's regions (Broekel, 2015) and efforts to incorporate standardisation into the research and innovation process (Zi & Blind, 2015). Russia's government stimulated a competitive university landscape in the 2010s which linked funding to scientific output and impact (Ivanov et al., 2016).

## 2.3. Triple and quadruple helix models

Increasing interdependence between industry, academia, and government has the advantage of optimising conditions for creative thinkers to work alongside innovators and entrepreneurs. In 1995, Etzkowitz and Leydesdorff described the collaborative relationship in knowledge based economic development between three stakeholders of innovation, namely universities, industry, and government, as the 'Triple helix' (Etzkowitz & Leydesdorff, 1995), extending the notion of interdependence from biology. Progressively integrated models see partnership between independent but separate actors gradually transform into a system of overlapping entities generating tri-lateral networks and hybrid organisations facilitated by, but not controlled by government (Etzkowitz & Leydesdorff, 2000; Leydesdorff & Etzkowitz, 1996).

According to Kang et al. (2019), Triple Helix theory was designed for established economies of Europe, the USA, and other western countries. Kang studied collaboration dynamics between stakeholders in China's two main science and technology innovation centres located in Shanghai and Beijing. The study and others attributed a strengthening cooperation between universities and industry to the shift in focus of China's universities from traditional research to an entrepreneurial university model (Kang et al., 2019; Zhu et al., 2022).

Other stakeholders may also enter the collaboration such as civil society, which extends the model to form the quadruple helix (Arnkil et al., 2010; Carayannis & Rakhmatullin, 2014), and the environment, which makes up the quintuple helix. The helix models are not evaluated in this study but may interact with and influence conclusions drawn from studies of societal stakeholder collaboration with academia (Carayannis et al., 2020).

Early comparisons of triple helix activity between countries used combined search terms for the local language translations of the words university, industry, and government (Park et al., 2005). Macro-level quantitative analyses may offer useful information, but public-private co-publication indicators need to be valid, reliable, and robust in order to form the basis for policy decision making (Tijssen, 2012).

## 3. Data and methods

### 3.1. Data

We sourced scholarly publications from the Dimensions database because of its broad coverage (Hook et al., 2018) and because it may index some relevant content not covered by other bibliometric databases (Paez, 2017; Visser et al., 2021). Specifically, we used the Dimensions database hosted by CWTS at Leiden University and focused on records published between 2013 and 2022 (10 full years).

The affiliations in Dimensions are mapped on to the Research Organization Registry (ROR) (https://ror.org/registry/), which defines more than 100,000 research organisations by type. Every organisation is manually assigned to an organisation type as described in Table 1.

Table 1. Organisation types in the Research Organization Registry

| Organisation type | Description |
| --- | --- |
| Education | A university or similar institution involved in providing education and educating/employing researchers. |
| Healthcare | A medical care facility such as hospital or medical clinic. Excludes medical schools, which should be categorized as "Education". |
| Company | A private for-profit corporate entity involved in conducting or sponsoring research. |
| Archive | An organisation involved in stewarding research and cultural heritage materials. Includes libraries, museums, and zoos. |
| Nonprofit | A non-profit and non-governmental organisation involved in conducting or funding research. |
| Government | An organisation that is part of or operated by a national or regional government and that conducts or supports research. |
| Facility | A specialized facility where research takes place, such as a laboratory or telescope or dedicated research area. |
| Other | Any organisation that does not fit the categories above. |

Source: ROR Data Structure

## 3.2. Methods

In order to study the collaboration between stakeholder types, we analysed the author affiliations on the papers in our dataset. In this study, a paper was linked to a stakeholder type only if a minimum proportion of its author affiliations were linked to that stakeholder type. In the first analysis, we tested the effect of setting different thresholds for counting the paper as being authored by a given stakeholder type. We considered four thresholds: 0% (one single author affiliation), 10%, 20%, and 30% of the author affiliations in a paper. For the rest of the study, we set the minimum threshold at 20% of the author affiliations. For instance, if a paper had 10 author affiliations, then it only counted as an academic paper if at least two of the author affiliations were with academic institutions.

As the vast majority of published research papers feature academic affiliations, we began our study by retrieving all papers in which at least 20% of the author affiliations were linked to academic institutions. This was defined as the overall academic dataset. Within the academic dataset, we then identified all those papers that featured collaboration with another stakeholder type (i.e. academia – industry, academia – nonprofit, or academia – government). If a paper features a minimum 20% author affiliations with academic institutions and a minimum 20% affiliations with industry, then it counts as academia – industry collaboration. Conversely, in the case that a paper with 25 author affiliations featured only one industry affiliation, it would not be counted as an industry collaboration because the industry contribution is not sufficient to meet the 20% affiliation threshold. For academic collaboration with each of the other stakeholder types, we computed the number and proportion of collaborative papers for each year in the period 2013-2022.

We then conducted further analyses of academic collaboration with other stakeholder types by country. We focused on the 25 countries with most academic publications along with the world and European Union as benchmarks (Table 2). In the country analysis, we based the country only on the academic affiliations. For instance, if a paper features academic affiliations from USA and China, and industry affiliations from India and Brazil, then it would only count as a paper for the USA and China. This enabled us to view international collaboration from the perspective of the academic community in a given country.

Table 2. Countries and their abbreviations

| Country/territory | Abbr. | Country/territory | Abbr. |
|---|---|---|---|
| United States | US | Russia | RU |

| European Union | EU | France | FR |
| --- | --- | --- | --- |
| China | CN | Iran | IR |
| United Kingdom | UK | Indonesia | ID |
| Japan | JP | Turkey | TR |
| Germany | DE | Netherlands | NL |
| India | IN | Poland | PL |
| Canada | CA | Sweden | SE |
| Italy | IT | Taiwan | TW |
| Australia | AU | Switzerland | CH |
| Brazil | BR | Belgium | BE |
| South Korea | KR | Malaysia | MY |
| Spain | ES | Denmark | DK |

Similarly, we conducted a collaboration analysis for each of the stakeholder pairs (academia – industry, academia – nonprofit, and academia – government) based on the field of research of the paper (Table 3).

Table 3. Fields of research and their abbreviations

| Field of Research | Abbr. | Field of Research | Abbr. |
| --- | --- | --- | --- |
| Agricultural, Veterinary & Food Sciences | AG | Environmental Sciences | EV |
| Biological Sciences | BS | Health Sciences | HS |
| Biomedical & Clinical Sciences | BC | History, Heritage & Archaeology | HH |

| Built Environment & Design | BE | Human Society | HU |
| --- | --- | --- | --- |
| Chemical Sciences | CS | Information & Computing Sciences | IC |
| Commerce, Management, Tourism & Services | CM | Language, Communication & Culture | LC |
| Creative Arts & Writing | CA | Law & Legal Studies | LL |
| Earth Sciences | ES | Mathematical Sciences | MS |
| Economics | EC | Philosophy & Religious Studies | PR |
| Education | ED | Physical Sciences | PS |
| Engineering | EN | Psychology | PY |

For academia – industry collaborative papers, we conducted an additional analysis to identify the companies most frequently involved in the collaboration for selected countries. We selected the two countries with steepest growth in relative academia – industry papers, and the two countries with steepest decline (provided the share of academia – industry papers exceeded 1% of the overall academic output for that country in 2022). We then identified the eight companies most frequently represented on the papers coauthored with academia in 2013 and the eight companies most frequently represented in 2022.

## 4. Results

### 4.1. Minimum author affiliation threshold

Stakeholder participation in collaborative papers was highest when no threshold was introduced, i.e. when one single author affiliation was sufficient to count a paper as belonging to a given stakeholder type (Figure 1). When 10% author affiliations were required to count the paper as a collaboration the share of collaborations began to drop. As the threshold of author affiliations for counting papers as collaborations grew, the share of collaborations dropped as expected. All results presented in the remainder of the study are based on a threshold of 20% author affiliations.

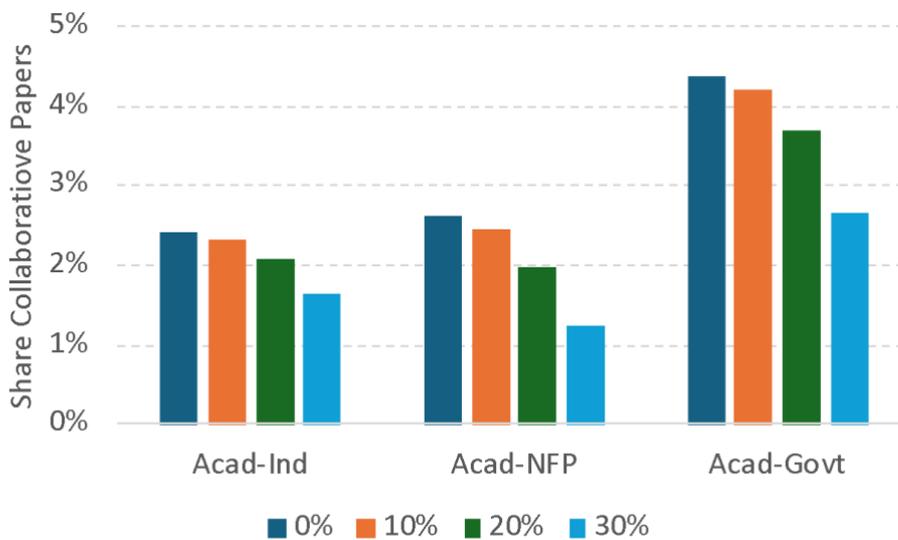

Figure 1. Share of authors to define stakeholder participation

## 4.2. Collaboration between stakeholders

In Figure 2, we show the number of papers that feature affiliations from both academic and industry organisations as a bar chart. The line graph in the same figure represents the collaborative papers as a proportion of the overall number of academic papers. While the number of collaborative papers has grown nearly 40% over the past 10 years, the share of academic papers with industry participation has dropped by 16%. This can be explained by the even greater growth in academic output that does not feature industry collaboration.

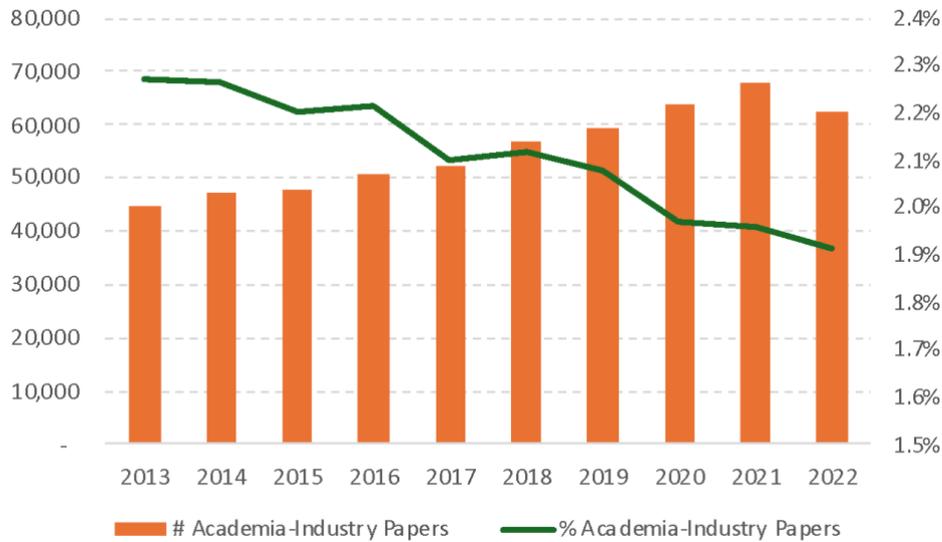

Figure 2. Academia - industry collaboration

In Figure 3, we show the same time series for academic papers that feature co-authors from nonprofit organisations. We see a similar pattern to the academia - industry collaboration described in figure 2. The number of collaborative papers between academia and nonprofit (figure 3) has grown almost 60% over the past 10 years, while the share of academic papers with nonprofit participation has dropped slightly. This means the number of collaborative papers between academia and nonprofit sector has grown slightly less than the overall academic output.

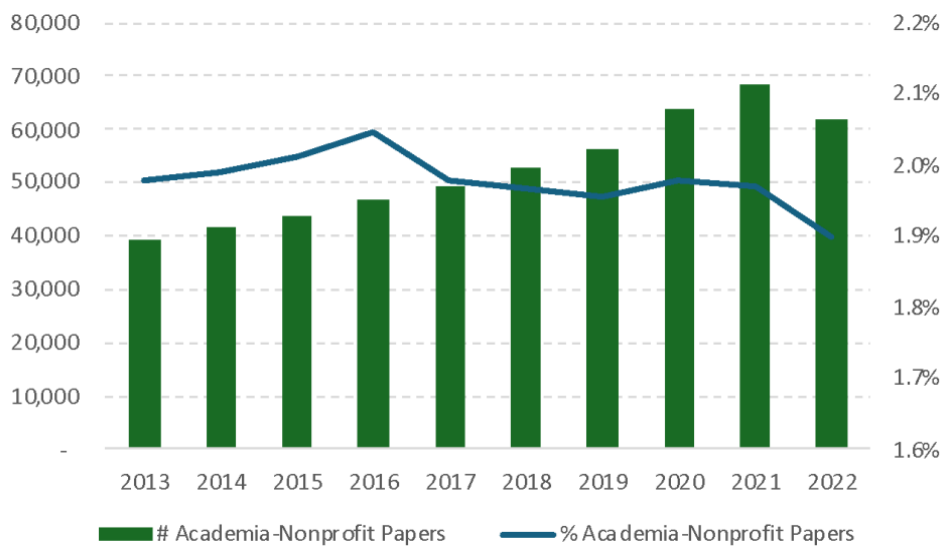

Figure 3. Academia - nonprofit collaboration

In Figure 4, we show the same time series for academic papers that feature co-authors from government organisations. Here we see that collaborative papers between academia and government have grown at almost 70% over the past 10 years. That growth has resulted in a 3% increase in the proportion of collaborative papers relative to the overall academic output.

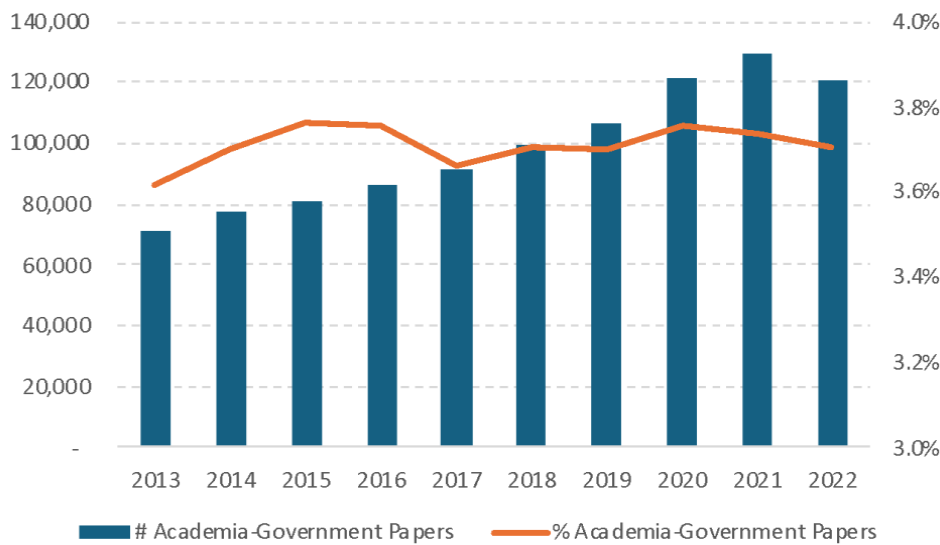

Figure 4. Academia - government collaboration

## 4.3. Country trends

In the next part of the analysis, we determined the 10-year collaboration trend for the 25 most productive countries along with the world and EU trends as benchmarks. For each country or territory, we determined the number of collaborative papers between academia and one of the other stakeholders as a proportion of the academia affiliated papers. In Figure 5, we show the share of academia – industry collaborative papers in 2013 and then again in 2022. A dot above the diagonal signifies an increase in share of academia – industry collaboration in that country or territory over the 10-year time period, while dots below it signify a decrease. China and Belgium, an outlier within the EU, showed substantial increases in academia – industry collaboration over the 10 years, while Australia, Taiwan, and Turkey appear to have maintained roughly the same share as 10 years earlier. For most countries, however, the figure shows a decline in academia - industry collaboration, in line with the overall declining trend shown in Figure 2. We will provide a more detailed analysis of these findings in section 4.5.

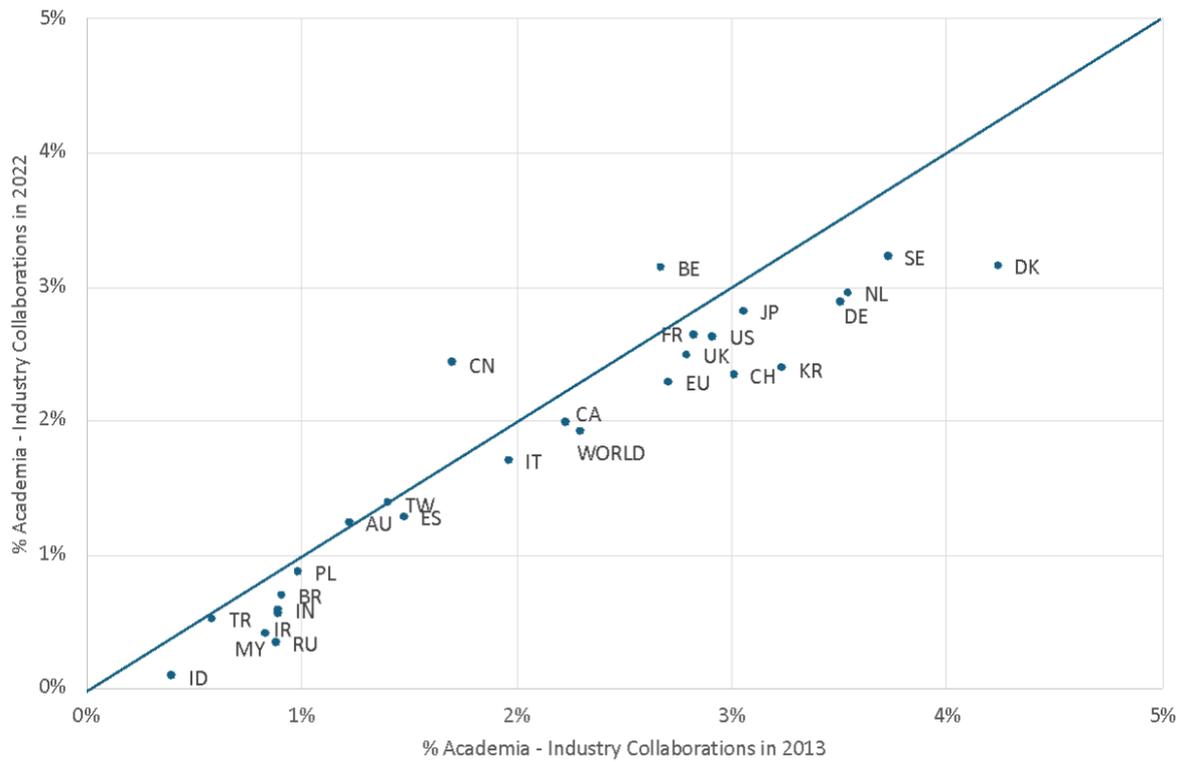

Figure 5. Academia - industry collaboration trend by country

Next, we present the 10-year trend for academia – nonprofit collaborations as a proportion of the overall academic output (Figure 6). Here we see the greatest gains in Spain, the United Kingdom, and Iran, with most of the other countries making modest growth. The most notable declines were in Taiwan, Indonesia, Belgium, and Russia.

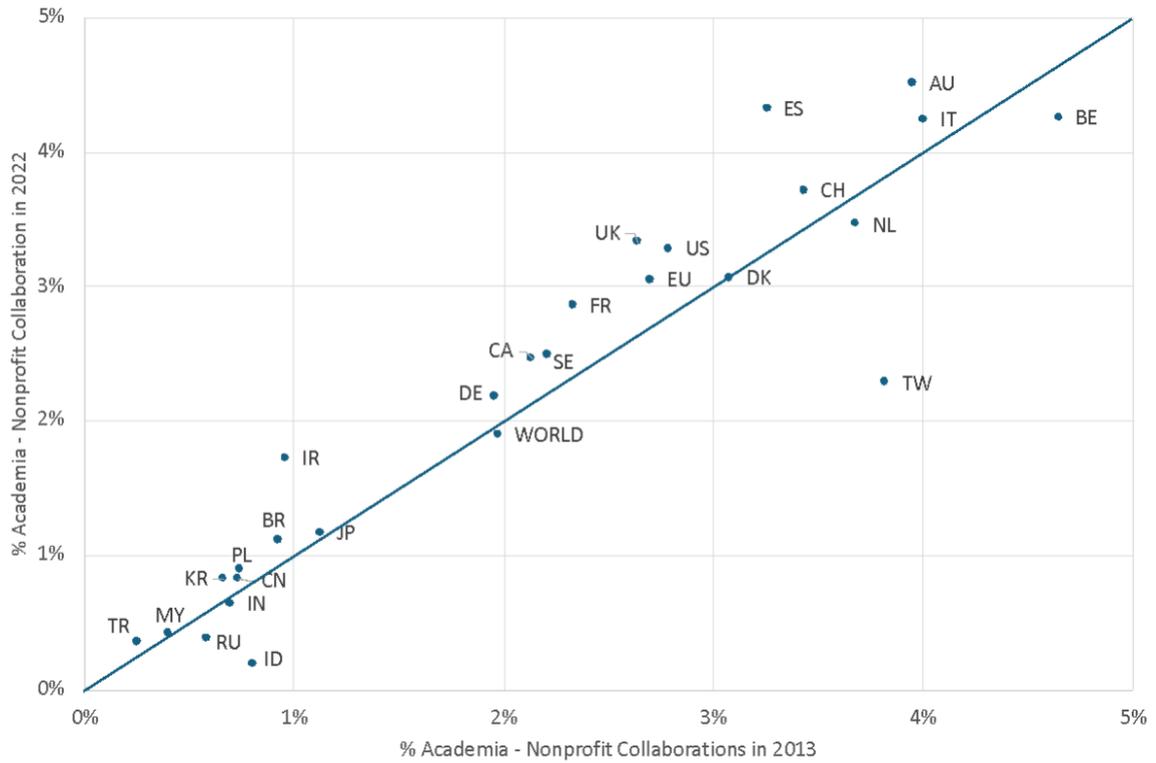

Figure 6. Academia - nonprofit collaboration trend by country

Next, we present the 10-year trend for academia – government collaborations as a proportion of the overall academic output (Figure 7). The picture for academia – government collaboration showed wide variance. The largest increases were seen in China, Malaysia, and the United Kingdom. The steepest declines were seen in France and Russia and Indonesia, Japan, Turkey, and Taiwan.

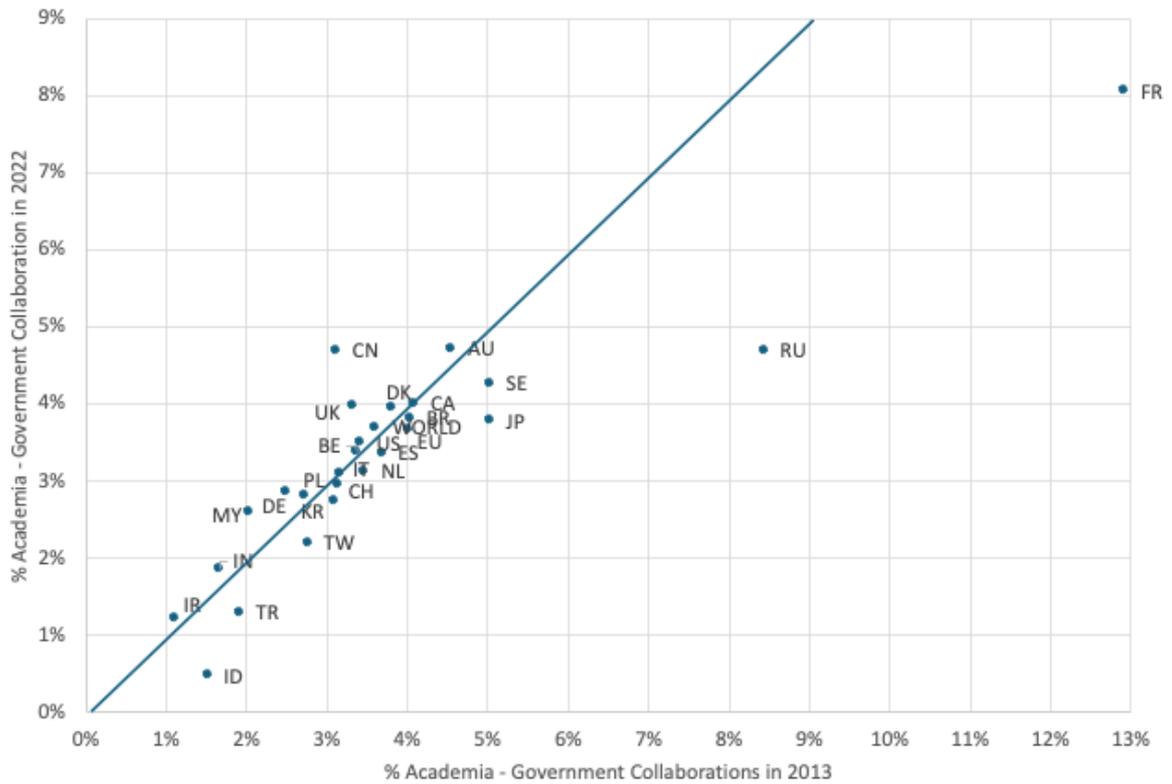

Figure 7. Academia - government collaboration trend by country

## 4.4. Field trends

We also analysed the trends of research for academic collaboration with each of the stakeholder types by subject field. In Figure 8, we see that the share of academia - industry collaborations with respect to overall academic output has fallen for all fields of research except earth sciences, which grew, and mathematical sciences, which retained almost the same share. Steeper declines were in the social sciences and humanities fields including creative arts & writing, law & legal studies, history, heritage, & archaeology. The science fields including engineering and the physical, biological, and chemical sciences declined less.

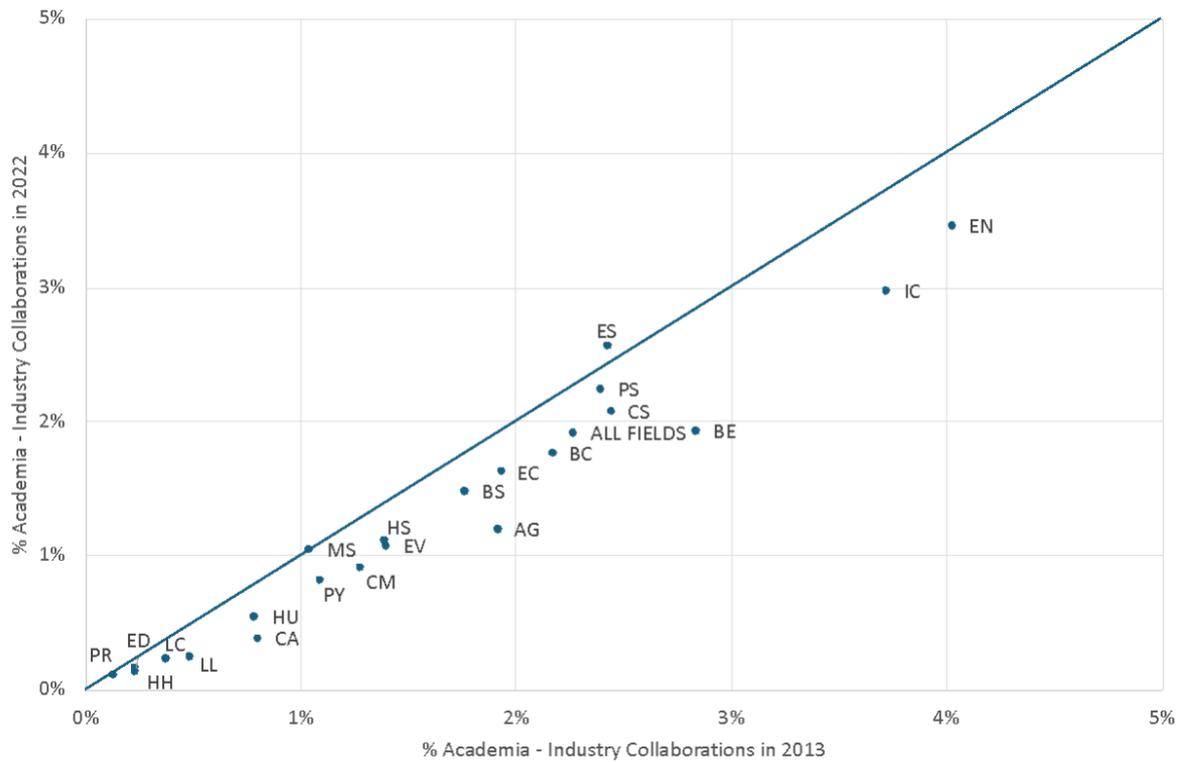

Figure 8. Academia - industry collaboration trend by field

The 10-year change in collaboration between academia and nonprofit organisations relative to the overall academic output is shown in Figure 9. Here the picture is rather diverse with marked increases in relative collaboration in philosophy & religious studies, history, heritage, & archaeology, and built environment & design. The steepest declines were noted in the creative arts & writing and the law & legal studies. The sciences tended to vary less and maintain a stable share of collaboration over the 10-year period.

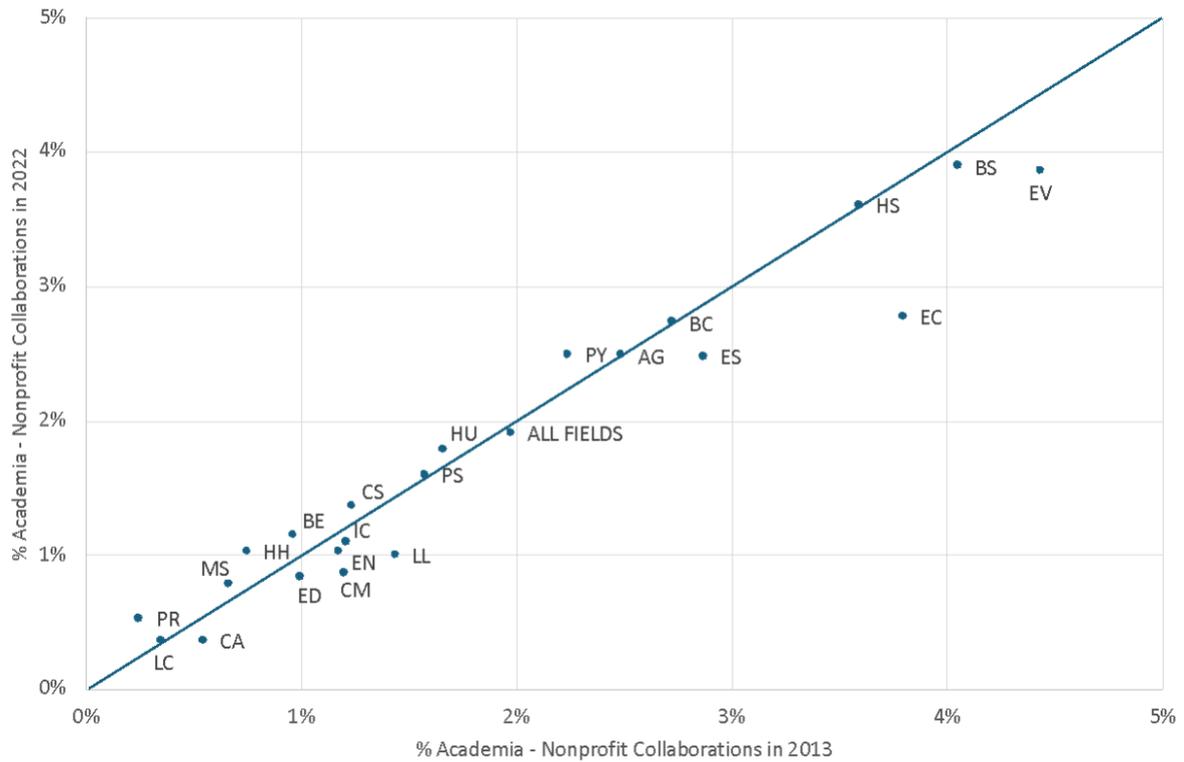

Figure 9. Academia - nonprofit collaboration trend by field

The collaboration between academia and government relative to the overall academic output is shown in Figure 10. In most fields, there was an increase in relative collaboration, most notably in the law & legal studies, philosophy & religious studies, built environment & design, history, heritage, & archaeology, and engineering. The steepest declines were seen in commerce, management, tourism, & services, economics, and environmental sciences. The sciences again maintained a relatively stable share.

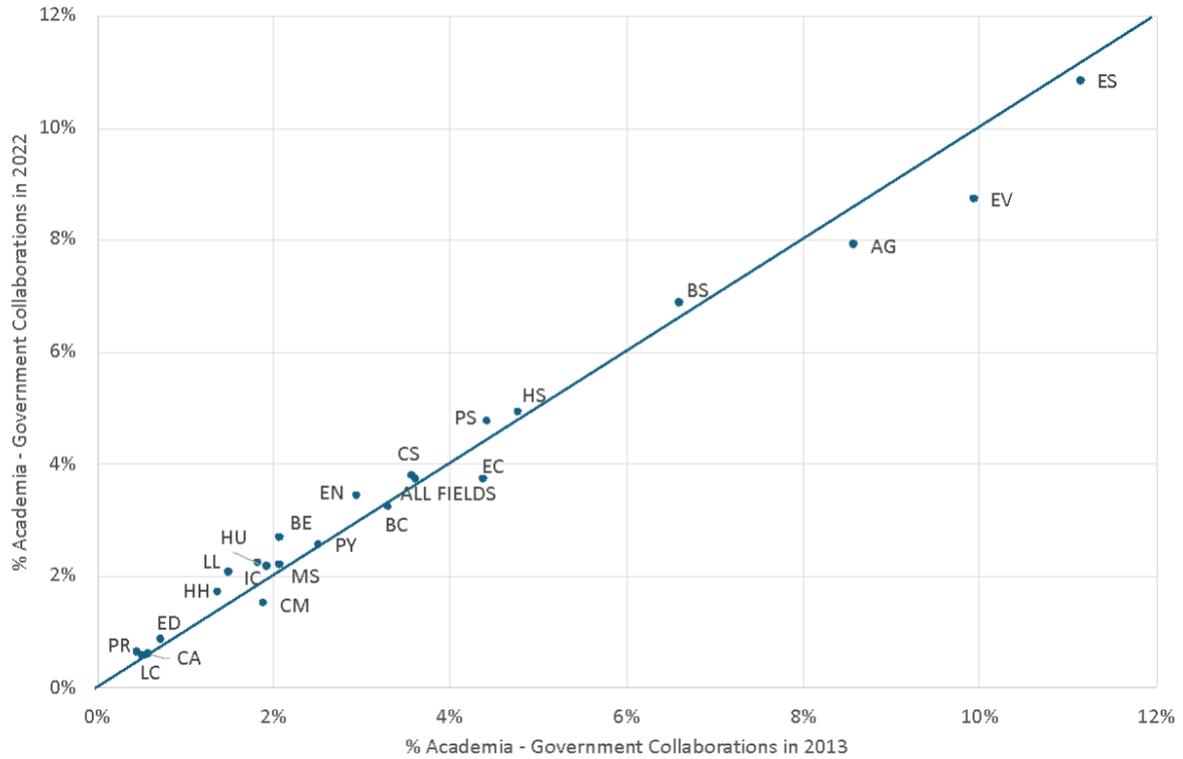

Figure 10. Academia - government collaboration trend by field

## 4.5. Industry partner analysis for selected countries

The findings presented in figures 3 and 4 show that academia – nonprofit and academia – government collaboration grew at roughly the same rate as overall academic output. However, figure 2 shows that academia – industry collaboration dropped by 16% relative to academic output. The latter was an interesting and surprising result in the context of the frequent calls on academia to close the gap with industry. The country-level findings presented in figure 5 could potentially shed light on whether the relative decline is a global one or if regional differences play a role. We therefore conducted an additional analysis of the academia – industry collaborations to identify the eight companies with which academia coauthored the most papers in the two countries with steepest growth (China and Belgium) and in the two countries with the steepest decline (South Korea and Switzerland) (table 4).

Table 4. Industry partners in academic collaboration: China and Belgium

| China 2013 | | China 2022 | |
|---|---|---|---|
| Company name | Papers | Company name | Papers |
| State Grid Corporation of China (China) | 682 | State Grid Corporation of China (China) | 3295 |
| China National Petroleum Corporation (China) | 242 | Sinopec (China) | 961 |
| Sinopec (China) | 194 | China National Petroleum Corporation (China) | 875 |
| Microsoft Research Asia (China) | 183 | China Electronics Technology Group Corporation (China) | 754 |
| China Electronics Technology Group Corporation (China) | 149 | China Shipbuilding Industry Corporation (China) | 593 |
| China North Industries Group Corporation (China) | 132 | China North Industries Group Corporation (China) | 545 |
| Aviation Industry Corporation of China (China) | 127 | China Southern Power Grid (China) | 509 |
| China National Offshore Oil Corporation (China) | 115 | Aviation Industry Corporation of China (China) | 463 |
| **Belgium 2013** | | **Belgium 2022** | |
| Company name | Papers | Company name | Papers |
| Janssen (Belgium) | 50 | Flanders Make (Belgium) | 175 |
| Siemens (Belgium) | 14 | Sciensano (Belgium) | 153 |
| GlaxoSmithKline (Belgium) | 12 | Janssen (Belgium) | 41 |
| UCB Pharma (Belgium) | 12 | Sanofi (United States) | 19 |
| Novartis (Switzerland) | 11 | Siemens (Belgium) | 15 |

| Bekaert (Belgium) | 9 | Regeneron (United States) | 12 |
| Holst Centre (Netherlands) | 9 | Sanofi (France) | 11 |
| Novartis (United States) | 8 | Icometrix (Belgium) | 10 |

In China, the top eight collaborating companies in 2022 coauthored more than four times the number of collaborations as the top eight companies in 2013. The growth appears to be generic as all the top collaborating companies in 2022 were already collaborating with academia in 2013. However, the number of co-authored papers has increased by several-fold over the 10-year time period. The biggest contributor to China's rise was by the State Grid Corporation of China, whose co-authored publication count grew nearly five-fold in the 10-year period studied.

In Belgium, collaboration with the top eight companies tripled between 2013 and 2022. The increased collaboration was largely due to over 300 collaborative papers with two companies, Flanders Make and Sciensano in 2022, with whom academia had no or negligible collaboration in 2013. Flanders Make supports the local manufacturing industry through innovative research projects to facilitate its customers' transition to industry 4.0. Flanders Make currently has open calls for research partnerships on strategic basic research projects such as 'Development and validation of a high-torque-dense actuator', and on industrial research valorisation and acceleration projects such as 'Situation-aware robust optimal vessel navigation and control' (Flanders Make, 2024). Sciensano is a healthcare company launched in 2018 that is also registered as a research institution under the Belgian Science Policy. The company aims to improve quality of life for humans and animals along with their shared environment in Belgium. Sciensano publishes details of projects on projects such as 'Research on PFAS contamination in the food chain', 'A citizen-driven crowdsourcing and feedback system to stimulate research and policy relating to Flemish and local food environments', and 'Development, testing, and implementation of the Belgian Patient Reported Experience measure for PAncreatic cancer caRE' (Sciensano, 2024).

For many countries there was a relative decline between 2013 and 2022 in collaboration between academia and industry (figure 5). The two countries showing the steepest decline were South Korea and Switzerland. In table 5, we present the top eight companies by number of publications co-authored with academic institutions in South Korea in 2013 and in 2022 followed by the corresponding companies collaborating with academic institutions in Switzerland.

Table 5. Industry partners in academic collaboration: South Korea and Switzerland

| South Korea 2013 | | South Korea 2022 | |
|---|---|---|---|
| Company name | Papers | Company name | Papers |
| Samsung (South Korea) | 730 | Samsung (South Korea) | 587 |
| LG Corporation (South Korea) | 155 | Hyundai Motors (South Korea) | 120 |
| Pohang Iron and Steel (South Korea) | 78 | LG Corporation (South Korea) | 112 |
| SK Group (South Korea) | 68 | SK Group (South Korea) | 97 |
| Korea Electric Power Corporation (South Korea) | 58 | Pohang Iron and Steel (South Korea) | 87 |
| Amorepacific (South Korea) | 51 | Hyundai Motor Group (South Korea) | 53 |
| Hyundai Motors (South Korea) | 32 | Korea Electric Power Corporation (South Korea) | 51 |
| CJ CheilJedang (South Korea) | 26 | Amorepacific (South Korea) | 32 |
| Switzerland 2013 | | Switzerland 2022 | |
| Company name | Papers | Company name | Papers |
| Novartis (Switzerland) | 51 | Roche (Switzerland) | 61 |

| Roche (Switzerland) | 31 | Novartis (Switzerland) | 54 |
| ABB (Switzerland) | 28 | Inspire | 24 |
| Walt Disney (Switzerland) | 26 | Roche (United States) | 21 |
| Nestlé (Switzerland) | 25 | Microsoft (United States) | 20 |
| Microsoft (United States) | 16 | Fluxim (Switzerland) | 15 |
| Siemens (Germany) | 16 | Novartis (United States) | 14 |
| Novartis (United States) | 11 | Philochem (Switzerland) | 14 |

In South Korea, the number of collaborative papers with top eight companies was slightly less in 2022 than it was in 2013. There was very little change in the companies participating in the collaborations, indeed the top seven companies in academic collaborations in 2013 were still among the top eight in 2022. The number of co-authored papers had dropped, which contributed to the overall decline in academic – industry collaboration in South Korea over the 10-year period studied. Notable declines were from Samsung (-20%) and LG Corporation (-28%).

In Switzerland, the number of collaborative papers with the top eight companies actually grew by 8% in absolute terms but declined relative to overall academic output. There were interesting changes in the identity of the leading companies collaborating with academia. From the list of top eight collaborating firms in 2013, ABB, Walt Disney, Nestlé, Microsoft, and Siemens all posted substantial declines in co-authored papers with academic institutions in Switzerland by 2022.

## 5. Discussion

### 5.1. Academia – Industry

We frequently hear research policymakers and higher education leaders citing the need for academia to work more closely with different societal stakeholders to maximise the benefits of research outcomes when addressing society's most pressing goals. Our results show that the number of academia – industry collaborative publications has consistently increased over time in absolute terms but has declined with respect to overall academic output. The idea that academic research in partnership with the private sector will help industry to innovate may have stimulated some collaborative research, but overall academic research output has grown faster without industry partnership than with it. Our findings show there is a wide variation in this trend depending on the country studied, with substantial relative growth in China and notable decline several countries including South Korea and Switzerland. In Europe, only Belgium has seen relative growth in academia – industry collaboration.

In 2019, China became the leading country in USPTO patent filings, with four Chinese universities among the top 10 education institutions in number of filed applications (WIPO, 2020). The impressive trend for growth in industry participation in China may have been influenced by the major government overhaul of laws and regulations surrounding research commercialisation dubbed China's Bayh-Dole Act (Huang et al., 2023; Yi & Long, 2021). In the past decade, universities have now assumed responsibility for implementing their own patents and sharing profits with researchers. This may be responsible for stimulating a spike in Chinese universities commercialising research. Industry funding of universities in China would then create conditions that could explain our reported increase in relative participation of industry in scientific research.

Two Belgian companies were responsible for three-quarters of the top eight companies academia – industry collaborations. Flanders Make supports local manufacturers and Sciensano is a commercial research organisation that focuses on improving local human and animal health. Academic collaboration with both these companies has started recently and both promote calls for collaborative research projects on their websites. The academic community has clearly responded to calls to work together with these two companies, and policymakers could potentially use the model as an example to encourage other similar collaborations.

Despite South Korea's government boost to stimulate academia – industry collaboration, researchers in the private sector still need to develop a culture of trust in their academic partners (Lee, 2014). Our reported decline in the relative number of academia – industry collaborative papers support previous work (e.g., Jones, 2024). In Switzerland, we did not find any obvious explanation for the decline in academia – industry collaboration.

### 5.2. Academia – Nonprofit

The absolute number of academia – nonprofit collaborative research papers has shown consistent year-on-year growth in the past 10 years that was almost in line with the steep growth in overall academic output. Most of the high-income countries showed relative gains in academia – nonprofit research collaboration, but the country with the steepest increase was Iran. In Iran, it has been argued that citizen engagement on civic societal issues will be among the emerging research themes and is a potential driver behind academia – nonprofit engagement (Farazmand et al., 2019).

### 5.3. Academia – Government

The absolute number of academia – government collaborative research papers shows consistent year-on-year growth in the past 10 years and is almost in line with the overall academic output. Countries with the most marked increase in academia – government collaboration were China, Malaysia, UK, Germany, India, and Iran. Meanwhile, the steepest declines were recorded in Russia, France, and Indonesia.

Relative academia – government collaboration declined on average in EU academic sectors, which is at odds with the EU Research and Innovation policy that aims to 'ensure technological breakthroughs are developed into viable products with real commercial potential – by building partnerships with industry and governments' (European Union, 2020). However, the EU's leading gain was recorded by Germany, whose most prolific government research institution in 2022 was the Helmholtz-Zentrum Hereon, a transdisciplinary centre that spends 100 million Euros annually on support for societal institutions through its expertise in tackling climate change, biomedicine, and new energy systems.

### 5.4. Limitations

In this study we have used three pairwise comparisons, each from the perspective of the academic sector. The main reason for this is because the vast majority of published research is by authors affiliated to academic institutions. Therefore, we used academic output as the denominator in our calculations and collaboration between academia and each of the other stakeholders as the numerator. So, the question we are addressing is about the extent to which academia has collaborated with other stakeholders rather than how much the stakeholders are collaborating with each other. We might be overlooking other scenarios such as industry providing contract research services to government, which would also count as transdisciplinary research but not be counted in this study.

The Dimensions database used for this study relies on the Research Organization Registry (ROR) classification system for defining research stakeholders. It is not always straightforward to classify institutions. For instance, the State Grid Corporation of China was classified as a 'Company', whereas it could be argued that it should be 'Government'. We also noted that the classification of some organisations was changed in ROR during the course of our study. The École Polytecnique Fédérale de Lausanne (EPFL) was initially classified as 'Facility', but later changed to 'Education' and 'Funder'. Its re-classification as 'Education' meant that papers with the EPFL affiliation were included in our study and counted as academia. However, these were isolated discrepancies and we considered them artefacts in the data rather than systematic errors. Other stakeholder types categorised by ROR, such as facilities, museums, and private hospitals, were not analysed in this study, but may be of interest, especially in studies of healthcare and studies of the arts and humanities. Despite these concerns, ROR routinely updates its classification, and we consider it the most appropriate classification system of research organisations for this study. Given the changing and sometimes heterogeneous nature of organisation types, we suggest some flexibility in interpreting the results.

One of the limitations of all bibliometric studies is that we can only analyse research outputs that are published. The academic sector is by far the most prominent contributor to published research among each of the stakeholders studied. This is expected because scholarly publication is one of the most important performance indicators for academics and for their universities (Dill & Soo, 2005; Hazelkorn, 2011; Marginson & van der Wende, 2007). University academics routinely publish in journals, books, and conferences, but other stakeholders may not.

Private sector companies for instance, often conduct research aiming to bring unique products or services to market and consequently make only a fraction of their findings public. This means that we do not know what share of industry research we are counting through scholarly publications. Similarly, government participation in major challenges is important because governments can focus a nation's attention and priority, and because only governments can set national agendas and make laws. Governments habitually publish reports, guidelines, or other types of documents that do not appear in bibliographic databases as scholarly works (Bickley et al., 2020). Scholarly publications from industry, government, and nonprofits are likely to represent only a fraction of the research these actors perform.

Introduction of a 10% minimum threshold of author affiliations to count a stakeholder as having a meaningful contribution to a study reduced the number of papers in our dataset. Raising that minimum threshold to 20% and then 30% accelerated the exclusion of papers. In order to allow our threshold to eliminate the most extreme cases of lopsided collaborative papers, without unduly reducing the dataset, we set the threshold at 20% for the remainder of the study. Although the choice of 20% is somewhat arbitrary, we considered it an improvement over having no threshold at all.

## 6. Conclusion

Research policymakers and academic leaders emphasise the benefits of closing the gap between academia and industry and collaborating more closely with society to address the most difficult challenges of our time. This bibliometric study of transdisciplinary research applies large-scale quantitative techniques to assess collaboration trends between academia on the one hand and industry, government, and nonprofit organisations on the other.

There appears to be a gap between the aspirations of research policymakers who are advocating closer collaboration between academia and other research stakeholders and what is happening in reality. The overall trends along with the picture in many countries and many fields is a relative decline in collaborative research with respect to the rapid growth in research by academia alone. Although collaboration is growing in absolute terms, from a relative point of view our results do not support the notion of increased collaboration, partnerships, or transdisciplinary research.

One possible general explanation for the relative decline is the current system of incentivising academic researchers on their scholarly publications in peer-reviewed journals. Perhaps many academics find it easier and quicker to publish with fellow academics, rather than through societal engagement projects. Collaboration with industry, nonprofits, or government may introduce partners with very different ideas of successful outcomes that do not involve published papers. Faced with conflicting pressures, academics might find the pressure to publish outweighs the need to conduct societal engagement.

We performed our analysis at global and at country level. This allowed us to observe wide variation in regional trends possibly influenced by changes in national research landscapes. A clear example is the opening of Chinese intellectual property laws, which has paved the way for universities to work more closely with industry on commercialisation of research. The resulting boost in academia – industry collaboration was unique to China according to our study.

The data used for this study were from the Dimensions database, which uses the organisation classification provided by the Research Organization Registry (ROR). The ROR organisation type categories are one way of defining stakeholders, but it is open for debate as to whether it is the right or best way. For the purposes of this paper, we have chosen to work with the ROR definitions used by Dimensions for pragmatic reasons and have revealed some interesting, high-level patterns. To drill into further detail, it may be useful to run follow-on studies that consider other data sources and stakeholder definitions.

Bibliometric studies generally look backwards but can reveal patterns that may then be linked to policy changes. Therefore, research policymakers will benefit from retrospective studies when setting out their new strategies on a range of policies such as funding, academic freedom, intellectual property management.

# References

Adams, Black, G. C., Clemmons, J. R., & Stephan, P. E. (2005). Scientific teams and institutional collaborations: Evidence from U.S. universities, 1981–1999. *Research Policy*, *34*(3), 259–285. https://doi.org/10.1016/j.respol.2005.01.014

Agarwal, R., & Ohyama, A. (2012). Industry or Academia, Basic or Applied? Career Choices and Earnings Trajectories of Scientists. *Management Science*, *59*(4), 950–970. https://doi.org/10.1287/mnsc.1120.1582

Arnkil, R., Järvensivu, A., Koski, P., & Piirainen, T. (2010). *Exploring Quadruple Helix Outlining user-oriented innovation models* (85; 2010). https://trepo.tuni.fi/bitstream/handle/10024/65758/978-951-44-8209-0.pdf

Auranen, O., & Nieminen, M. (2010). University research funding and publication performance—An international comparison. *Research Policy*, *39*(6), 822–834. https://doi.org/10.1016/j.respol.2010.03.003

Banal-Estañol, A., Jofre-Bonet, M., & Lawson, C. (2015). The double-edged sword of industry collaboration: Evidence from engineering academics in the UK. *Research Policy*, *44*(6), 1160–1175. https://doi.org/10.1016/j.respol.2015.02.006

Beaudry, C., & Allaoui, S. (2012). Impact of public and private research funding on scientific production: The case of nanotechnology. *Research Policy*, *41*(9), 1589–1606. https://doi.org/10.1016/j.respol.2012.03.022

Bickley, M. S., Kousha, K., & Thelwall, M. (2020). Can the impact of grey literature be assessed? An investigation of UK government publications cited by articles and books. *Scientometrics*, *125*(2), 1425–1444. https://doi.org/10.1007/s11192-020-03628-w

Bikard, M., Vakili, K., & Teodoridis, F. (2018). When Collaboration Bridges Institutions: The Impact of University–Industry Collaboration on Academic Productivity. *Organization Science*, *30*(2), 426–445. https://doi.org/10.1287/orsc.2018.1235

Bjerregaard, T. (2010). Industry and academia in convergence: Micro-institutional dimensions of R&D collaboration. *Technovation*, *30*(2), 100–108. https://doi.org/10.1016/j.technovation.2009.11.002

Brewer, G. D. (1999). The Challenges of Interdisciplinarity. *Policy Sciences*, *32*(4), 327–337. http://www.jstor.org/stable/4532473

Broekel, T. (2015). Do Cooperative Research and Development (R&D) Subsidies Stimulate Regional Innovation Efficiency? Evidence from Germany. *Regional Studies*, *49*(7), 1087–1110. https://doi.org/10.1080/00343404.2013.812781

Carayannis, Acikdilli, G., & Ziemnowicz, C. (2020). Creative Destruction in International Trade: Insights from the Quadruple and Quintuple Innovation Helix Models. *Journal of the Knowledge Economy*, *11*(4), 1489–1508. https://doi.org/10.1007/s13132-019-00599-z

Carayannis, Del Giudice, M., & Rosaria Della Peruta, M. (2014). Managing the intellectual capital within government-university-industry R&D partnerships. *Journal of Intellectual Capital*, *15*(4), 611–630. https://doi.org/10.1108/JIC-07-2014-0080

Carayannis, & Rakhmatullin, R. (2014). The Quadruple/Quintuple Innovation Helixes and Smart Specialisation Strategies for Sustainable and Inclusive Growth in Europe and Beyond. *Journal of the Knowledge Economy*, *5*(2), 212–239. https://doi.org/10.1007/s13132-014-0185-8

Cornforth, C., Hayes, J. P., & Vangen, S. (2014). Nonprofit–Public Collaborations: Understanding Governance Dynamics. *Nonprofit and Voluntary Sector Quarterly*, *44*(4), 775–795. https://doi.org/10.1177/0899764014532836

Czarnitzki, D., Grimpe, C., & Toole, A. A. (2015). Delay and secrecy: does industry sponsorship jeopardize disclosure of academic research? *Industrial and Corporate Change*, *24*(1), 251–279. https://doi.org/10.1093/icc/dtu011

Dill, D. D., & Soo, M. (2005). Academic quality, league tables, and public policy: A cross-national analysis of university ranking systems. *Higher Education*, *49*(4), 495–533. https://doi.org/10.1007/s10734-004-1746-8

ETH Zürich. (2024). *Collaborate with industry*. ETH Zürich. https://ethz.ch/en/industry/researchers/collaborations.html

Etzkowitz, H., & Leydesdorff, L. (1995). The Triple Helix--University-industry-government relations: A laboratory for knowledge based economic development. *EASST Review*, *14*(1), 14–19. https://papers.ssrn.com/sol3/papers.cfm?abstract_id=2480085

Etzkowitz, H., & Leydesdorff, L. (2000). The dynamics of innovation: from National Systems and "Mode 2" to a Triple Helix of university–industry–government relations. *Research Policy*, *29*(2), 109–123. https://doi.org/10.1016/S0048-7333(99)00055-4

European Union. (2020). *Leading innovation through EU research*. Research and Innovation - Priorities and Actions.

Eurostat. (2024). *GERD by sector of performance*. https://ec.europa.eu/eurostat/databrowser/view/RD_E_GERDREG/default/table?lang=en

Fang, L. H., Lerner, J., & Wu, C. (2017). Intellectual Property Rights Protection, Ownership, and Innovation: Evidence from China. *The Review of Financial Studies*, *30*(7), 2446–2477. https://doi.org/10.1093/rfs/hhx023

Farazmand, A., Danaeefard, H., Mostafazadeh, M., & Sadeghi, M. R. (2019). Trends in Public Administration Research: A Content Analysis of Iranian Journal Articles (2004-2017). *International Journal of Public Administration*, *42*(10), 867–879. https://doi.org/10.1080/01900692.2019.1598689

Flanders Make. (2024). *Our Research*. Research. https://www.flandersmake.be/en/offer/research

Gibbons, M., Limoges, C., Nowotny, H., Schwartzman, S., Scott, P., & Trow, M. (1994). *The New Production of Knowledge : The Dynamics of Science and Research in Contemporary Societies*. SAGE Publications Ltd. http://digital.casalini.it/9781446265871

Glied, S., Wittenberg, R., & Israeli, A. (2018). Research in government and academia: the case of health policy. *Israel Journal of Health Policy Research*, *7*(1), 35. https://doi.org/10.1186/s13584-018-0230-3

Goldfarb, B. (2008). The effect of government contracting on academic research: Does the source of funding affect scientific output? *Research Policy*, *37*(1), 41–58. https://doi.org/10.1016/j.respol.2007.07.011

Greenhalgh, T., Jackson, C., Shaw, S., & Janamian, T. (2016). Achieving Research Impact Through Co-creation in Community-Based Health Services: Literature Review and Case Study. *The Milbank Quarterly*, *94*(2), 392–429. https://doi.org/10.1111/1468-0009.12197

Haddon, C., & Sasse, T. (2019). *How academia can work with government*. https://apo.org.au/sites/default/files/resource-files/2019-04/apo-nid245496.pdf

Hazelkorn, E. (2011). Rankings and the reshaping of higher education: The battle for world-class excellence. In *Rankings and the Reshaping of Higher Education: The Battle for World-Class Excellence*. Palgrave Macmillan. https://doi.org/10.1057/9780230306394

Hernandez-Aguilar, C., Dominguez-Pacheco, A., Martínez-Ortiz, E. J., Ivanov, R., López Bonilla, J. L., Cruz-Orea, A., & Ordonez-Miranda, J. (2020). Evolution and characteristics of the transdisciplinary perspective in research: a literature review. *Transdisciplinary Journal of Engineering & Science*, *11*, 158–188. https://doi.org/10.22545/2020/00140

Hirsch Hadorn, G., Bradley, D., Pohl, C., Rist, S., & Wiesmann, U. (2006). Implications of transdisciplinarity for sustainability research. *Ecological Economics*, *60*(1), 119–128. https://doi.org/10.1016/j.ecolecon.2005.12.002

Hook, D. W., Porter, S. J., & Herzog, C. (2018). Dimensions: Building Context for Search and Evaluation. *Frontiers in Research Metrics and Analytics*, *3*, 23. https://doi.org/10.3389/frma.2018.00023

Huang, C., Cao, C., & Coreynen, W. (2023). Stronger and more just? Recent reforms of China's intellectual property rights system and their implications. *Asia Pacific Journal of Innovation and Entrepreneurship*, *ahead-of-p*(ahead-of-print). https://doi.org/10.1108/APJIE-04-2023-0081

Ivanov, V. V, Markusova, V. A., & Mindeli, L. E. (2016). Government investments and the publishing activity of higher educational institutions: Bibliometric analysis. *Herald of the Russian Academy of Sciences*, *86*(4), 314–321. https://doi.org/10.1134/S1019331616040031

Jones, R. S. (2024). Improving Korea's Innovation System. *The Peninsula*. https://keia.org/the-peninsula/improving-koreas-innovation-system/

Kang, W., Zhao, S., Song, W., & Zhuang, T. (2019). Triple helix in the science and technology innovation centers of China from the perspective of mutual information: a comparative study between Beijing and Shanghai. *Scientometrics*, *118*(3), 921–940. https://doi.org/10.1007/s11192-019-03017-y

Kim, M. (2016). The Relationship of Nonprofits' Financial Health to Program Outcomes: Empirical Evidence From Nonprofit Arts Organizations. *Nonprofit and Voluntary Sector Quarterly*, *46*(3), 525–548. https://doi.org/10.1177/0899764016662914

Lee. (2014). University–Industry R&D Collaboration in Korea's National Innovation System. *Science, Technology and Society*, *19*(1), 1–25. https://doi.org/10.1177/0971721813514262

Lee, & Nowell, B. (2014). A Framework for Assessing the Performance of Nonprofit Organizations. *American Journal of Evaluation*, *36*(3), 299–319. https://doi.org/10.1177/1098214014545828

Lee, Y. S. (2000). The Sustainability of University-Industry Research Collaboration: An Empirical Assessment. *The Journal of Technology Transfer*, *25*(2), 111–133. https://doi.org/10.1023/A:1007895322042

Leydesdorff, L. (2003). The mutual information of university-industry-government relations: An indicator of the Triple Helix dynamics. *Scientometrics*, *58*(2), 445–467. https://doi.org/10.1023/A:1026253130577

Leydesdorff, L., & Etzkowitz, H. (1996). Emergence of a Triple Helix of university-industry-government relations. *Science and Public Policy*, *23*(5), 279–286. https://doi.org/10.1093/spp/23.5.279

Manjarrés-Henríquez, L., Gutiérrez-Gracia, A., Carrión-García, A., & Vega-Jurado, J. (2009). The Effects of University–Industry Relationships and Academic Research On Scientific Performance: Synergy or Substitution? *Research in Higher Education*, *50*(8), 795–811. https://doi.org/10.1007/s11162-009-9142-y

Marginson, S., & van der Wende, M. (2007). To Rank or To Be Ranked: The Impact of Global Rankings in Higher Education. *Journal of Studies in International Education*, *11*(3–4), 306–329. https://doi.org/10.1177/1028315307303544

Marijan, D., & Sen, S. (2022). Industry–Academia Research Collaboration and Knowledge Co-Creation: Patterns and Anti-Patterns. *ACM Trans. Softw. Eng. Methodol.*, *31*(3). https://doi.org/10.1145/3494519

Mauser, W., Klepper, G., Rice, M., Schmalzbauer, B. S., Hackmann, H., Leemans, R., & Moore, H. (2013). Transdisciplinary global change research: the co-creation of knowledge for sustainability. *Current Opinion in Environmental Sustainability*, *5*(3), 420–431. https://doi.org/10.1016/j.cosust.2013.07.001

National Institutes of Health. (2012). NIH launches collaborative program with industry and researchers to spur therapeutic development. *News Releases*. https://www.nih.gov/news-events/news-releases/nih-launches-collaborative-program-industry-researchers-spur-therapeutic-development

National Institutes of Health. (2013). NIH to fund collaborations with industry to identify new uses for existing compounds. *News Releases*. https://www.nih.gov/news-events/news-releases/nih-fund-collaborations-industry-identify-new-uses-existing-compounds

Paez, A. (2017). Gray literature: An important resource in systematic reviews. *Journal of Evidence-Based Medicine*, *10*(3), 233–240. https://doi.org/10.1111/jebm.12266

Park, H. W., Hong, H. D., & Leydesdorff, L. (2005). A comparison of the knowledge-based innovation systems in the economies of South Korea and the Netherlands using Triple Helix indicators. *Scientometrics*, *65*(1), 3–27. https://doi.org/10.1007/s11192-005-0257-4

Perkmann, M., & Walsh, K. (2009). The two faces of collaboration: impacts of university-industry relations on public research. *Industrial and Corporate Change*, *18*(6), 1033–1065. https://doi.org/10.1093/icc/dtp015

Polman, P., & Winston, A. S. (2021). *Net positive* (Marrathon). Harvard Business Review Press.

Ponds, R., van Oort, F., & Frenken, K. (2007). The geographical and institutional proximity of research collaboration. *Papers in Regional Science*, *86*(3), 423 – 443. https://doi.org/10.1111/j.1435-5957.2007.00126.x

Rey-Garcia, M., Liket, K., Alvarez-Gonzalez, L. I., & Maas, K. (2017). Back to Basics. *Nonprofit Management and Leadership*, *27*(4), 493–511. https://doi.org/10.1002/nml.21259

Roach, M., & Sauermann, H. (2010). A taste for science? PhD scientists' academic orientation and self-selection into research careers in industry. *Research Policy*, *39*(3), 422–434. https://doi.org/10.1016/j.respol.2010.01.004

Savage, N. (2017). Industry links boost research output. *Nature Index*. https://www.nature.com/articles/d41586-017-07422-2

Sciensano. (2024). *Projects*. https://www.sciensano.be/en/projects

Seidl, R., Brand, F. S., Stauffacher, M., Krütli, P., Le, Q. B., Spörri, A., Meylan, G., Moser, C., González, M. B., & Scholz, R. W. (2013). Science with Society in the Anthropocene. *AMBIO*, *42*(1), 5–12. https://doi.org/10.1007/s13280-012-0363-5

Stanford University. (2024). *Industry Collaborations*. Stanford University, Engineering. https://engineering.stanford.edu/get-involved/industry-collaborations

The Russell Group of Universities. (2023). *Russell Group welcomes Chancellor prioritising innovation to drive growth and productivity in Autumn Statement*. Russell Group.

Thompson, M. A., Owen, S., Lindsay, J. M., Leonard, G. S., & Cronin, S. J. (2017). Scientist and stakeholder perspectives of transdisciplinary research: Early attitudes, expectations, and tensions. *Environmental Science & Policy*, *74*, 30–39. https://doi.org/10.1016/j.envsci.2017.04.006

Tijssen, R. J. W. (2012). Co-authored research publications and strategic analysis of public–private collaboration. *Research Evaluation*, *21*(3), 204–215. https://doi.org/10.1093/reseval/rvs013

Tijssen, R. J. W., van Leeuwen, T. N., & van Wijk, E. (2009). Benchmarking university-industry research cooperation worldwide: performance measurements and indicators based on co-authorship data for the world's largest universities. *Research Evaluation*, *18*(1), 13–24. https://doi.org/10.3152/095820209X393145

Tijssen, R. J. W., Yegros-Yegros, A., & Winnink, J. J. (2016). University–industry R&D linkage metrics: validity and applicability in world university rankings. *Scientometrics*, *109*(2), 677 – 696. https://doi.org/10.1007/s11192-016-2098-8

Tsinghua University. (2024). *Collaborating Institutions*. Research. https://www.tsinghua.edu.cn/en/Research/Collaborating_Institutions.htm

United Nations General Assembly. (2015). *Transforming our world: the 2030 Agenda for Sustainable Development*. Resolution Adopted by the General Assembly on 25 September 2015. https://documents-dds-ny.un.org/doc/UNDOC/GEN/N15/291/89/PDF/N1529189.pdf?OpenElement

Visser, M., van Eck, N., & Waltman, L. (2021). Large-scale comparison of bibliographic data sources: Scopus, Web of Science, Dimensions, Crossref, and Microsoft Academic. *Quantitative Science Studies*, *2*(1), 20–41. https://doi.org/10.1162/qss_a_00112

WIPO. (2020). *China Becomes Top Filer of International Patents in 2019 Amid Robust Growth for WIPO's IP Services, Treaties and Finances*. Press Release. https://www.wipo.int/pressroom/en/articles/2020/article_0005.html

Yi, W., & Long, C. X. (2021). Does the Chinese version of Bayh-Dole Act promote university innovation? *China Economic Quarterly International*, *1*(3), 244–257. https://doi.org/10.1016/j.ceqi.2021.09.003

Zhang, Patton, D., & Kenney, M. (2013). Building global-class universities: Assessing the impact of the 985 Project. *Research Policy*, *42*(3), 765–775. https://doi.org/10.1016/j.respol.2012.10.003

Zhang, Y., & Zou, Z. (Stephen). (2022). From academy to industry: China's new trend and policies on academic technology transfer. *IAM Life Sciences*. https://www.iam-media.com/global-guide/global-life-sciences/2022/article/academy-industry-chinas-new-trend-and-policies-academic-technology-transfer

Zhou, P., Tijssen, R., & Leydesdorff, L. (2016). University-Industry Collaboration in China and the USA: A Bibliometric Comparison. *PLOS ONE*, *11*(11), e0165277. https://doi.org/10.1371/journal.pone.0165277

Zhou, & Wang, M. (2023). The role of government-industry-academia partnership in business incubation: Evidence from new R&D institutions in China. *Technology in Society*, *72*, 102194. https://doi.org/10.1016/j.techsoc.2022.102194

Zhu, Z., Cui, S., Wang, Y., & Zhu, Z. (2022). Exploration on the collaborative relationship between government, industry, and university from the perspective of collaborative innovation. *Applied Mathematics and Nonlinear Sciences*, *7*(2), 903–912. https://doi.org/10.2478/amns.2021.2.00174

Zi, A., & Blind, K. (2015). Researchers' participation in standardisation: a case study from a public research institute in Germany. *The Journal of Technology Transfer*, *40*(2), 346–360. https://doi.org/10.1007/s10961-014-9370-y

## Acknowledgements

Ton van Raan and Ludo Waltman for expert advice and guidance throughout the study. Robert Tijssen for feedback on an earlier version of the paper.

## Competing interests